\documentclass[showpacs,preprintnumbers,amsmath,amssymb,11pt]{revtex4}
\usepackage{graphicx}
\usepackage{dcolumn}
\usepackage{bm}

\begin{document}

\title{Relativistic Theory of the Electric Dipole Moment of an Atom due to the Electric Dipole Moment of an Electron }

\author{Debashis Mukherjee \protect \footnote[2] {E-mail: pcdm@iacs.res.in}}
\affiliation{Raman Center for Atomic, Molecular and Optical Sciences, IACS, Kolkata 70032, India}
\author{B. K. Sahoo}
\affiliation{KVI, University of Groningen, NL-9747 AA Groningen, The Netherlands}
\author{H. S. Nataraj and B. P. Das}
\affiliation{Non-accelerator Particle Physics Group, Indian Institute of Astrophysics, Bangalore-560034, India}
\date{Received date; Accepted date}

\begin{abstract}
\noindent
The relativistic theory for the electric dipole moment (EDM) of paramagnetic 
atoms arising from the electric dipole moment of the electron is presented. A novel approach using the relativistic 
coupled-cluster method that incorporates the residual Coulomb interaction to all orders
and a weak parity and time-reversal violating interaction to one order has been employed in Fr to obtain
the enhancement of the  EDM of that atom compared to the EDM of the electron.
Trends of the different correlation effects and leading contributions from
different physical states are discussed. 
Our result in combination with that of the Fr EDM that is currently in progress, has the potential to probe the validity of the Standard Model (SM) of elementary particle physics. 
\end{abstract} 
\maketitle

\section{Introduction}
The Standard Model (SM) of elementary particle physics given by Glashow, 
Weinberg and Salam described by the $SU(3) \times SU(2) \times U(1)$ gauge 
group satisfactorily explains most of the phenomena observed so far on the fundamental interactions 
\cite{glashow,weinberg,salam}. However, this model is widely believed to be 
incomplete.  
Indeed a question of great importance is  whether there is any new physics beyond
the highly celebrated SM of elementary particles. Any kind of experimental
evidence supporting the extensions of the SM like the left-right symmetric model \cite{mohapatra}, the
multi-Higgs model \cite{barr} 
or supersymmetry models 
\cite{liu} will have a
significant impact on our understanding of the fundamental forces between the elementary particles in nature. 

  Detailed studies of the fundamental interactions have a bearing on chemical and biological processes as well. For example, the violation of parity is
believed to be
one of the fundamental reasons behind the "homochirality" of bio-molecules
\cite{cline}. In the same
vein, any time reversal symmetry ($\mathcal{T}$) violation will lead to  a certain asymmetry
in chemical phenomena whose importance will depend on the actual
magnitude of the asymmetry observed in the experiments. A case in point is the emergence of enantiomorphic
 excess where none is expected from the electromagnetic interaction \cite{barron}. One can in fact, distinguish
between the chirality brought out by $\mathcal{T}$-invariant and $\mathcal{T}$-noninvariant enantiomorphism, introducing the concept of
"false chirality" in the latter situation. Clearly if the predictions
by the SM and those beyond, relating to certain symmetry violations are observed, they will have profound chemical
and biological implications. Additionally, if the predictions from the SM and some of its extensions differ, then this will lead to different
expected magnitudes of the effect. This may bring observational efficiency to the limit achievable currently. Fixing
the limit of observability then depends on the accuracy of electronic structure theories \cite{barron}. This is why studies on the predictions
as defined by the SM and beyond are of immediate interest to chemists.

The EDM ($D$) associated with the nondegenerate state of a physical system like an atom can be shown to be proportional to the angular momentum ($J$) of that 
on the basis of the Wigner-Eckart theorem \cite{sakurai}. 
It can be proved that the existence of a nonzero electric
dipole moment (EDM) which is the expectation value of operator $D$ in a given atomic
state would imply the violations of the $\mathcal{T}$ invariance and parity
($\mathcal{P}$) symmetries \cite{lee, sandars}. The SM which conserves the combined
transformations of charge conjugation ($\mathcal{C}$), $\mathcal{P}$ and $\mathcal{T}$ symmetries 
(known as the CPT theorem \cite{luders}), predicts 
$\mathcal{P}$ and $\mathcal{CP}$ violations in weak interactions. A possible 
observation of a 
nonzero EDM in an atom would be a direct signature of $\mathcal{T}$ 
violation. An EDM has not been observed so far for 
whether in elementary particles or composite
systems. The SM predicts an upper limit on the electron EDM of the order 
of $10^{-38}e-cm$ \cite{bernreuther,khriplovich}, while other suggested models predict this value almost
ten orders of magnitude larger. Therefore, measurements 
of EDM in atomic systems are necessary to test the SM. 
Thus, an
unambiguous observation of  a nonzero EDM of any fundamental particle
including the electron will undoubtedly unveil a new arena of physics
beyond the SM. A direct measurement
of the electron EDM is not possible using accelerator approaches due to the
charge of the electron, however it could be obtained by combining the results 
measurements and calculations  of atomic EDMs using 
the table-top experiments.

 In 1963, Schiff \cite{schiff} argued that the permanent EDM of an atom 
vanishes under the assumption that the constituents of non-relativistic  point-like systems 
interact only electrostatically. It was, however, put forth 
by Salpeter \cite{salpeter} and Sachs and Scwebel \cite{sachs} as early as 
1958-1959 that when relativistic effects are included it is possible for an 
EDM on the electron to give rise to an EDM on the atom as a whole. Later, it 
was 
rigorously proved by Sandars \cite{sandars} that by introducing the correct 
relativistic form of the interaction Hamiltonian into the Dirac equation in 
a Lorentz covariant manner, as emphasized by Salpeter, one will get non-zero 
EDM for the atom when an electron is assumed to have intrinsic EDM. 

  In principle, an atom being a composite many-body system will have many 
sources of intrinsic EDMs arising from its constituent particles and 
their ${\mathcal P, T}$ violating electron-nucleus interactions. At the elementary particle 
level; electrons may possess intrinsic EDMs,  describing by a coupling constant 
$d_e$ which may directly contribute to the atomic EDM. The electron EDM can also  
interact with the central electrostatic field of an atom and can produce 
an atomic EDM. The interactions between electrons - quarks may manifest at 
different levels as electrons - nucleons to electrons - nucleus 
interactions. The ${\mathcal P}$ and ${\mathcal T}$ violating electron-nucleus 
interactions can either be scalar--pseudo-scalar (S-PS), 
with a 
coupling constant $C_s$ or a tensor--pseudo-tensor (T-PT) interaction describing 
with a coupling constant $C_T$.

   The paramagnetic (open-shell) atoms will in general be sensitive to the 
contribution from the intrinsic EDM of the electrons and their ${\mathcal P, \; \text{and}\; T}$-violating S-PS interactions with the nucleus whereas, the 
EDM in diamagnetic (closed-shell) atoms will arise from the EDM of the nucleus
and their ${\mathcal P, \; \text{and}\; T}$-violating T-PT interactions  
with the electrons. This can be understood as the electron EDM is a spin 
dependent property, hence it contributes to open-shell atoms because 
of the unpaired valence electron. However, for a closed-shell atom because 
of the Fermi exclusion principle, it will add up to zero if hyperfine interactions are excluded.
These ${\mathcal P}$- and ${\mathcal T}$-violating electron-nucleus
interactions will provide a useful tool in understanding the ${\mathcal CP}$
violation from the semi-leptonic sectors. Using the knowledge of
${\mathcal CP}$-violation obtained in these atomic EDM experiments one can
also constrain different models of ${\mathcal CP}$-violation.

  It was shown by Sandars that by choosing suitable atoms and favorable 
electronic states one can get an enhanced EDM for an atom which may even be 
a few orders of magnitude larger than that of the free electron 
\cite{sandars1}. In particular, he demonstrated by carrying out relativistic
calculations that the atomic EDMs of thallium (Tl) and caesium (Cs) are two orders of
magnitude
larger than the EDM of the electron. It was this important result which
provided
the impetus for the first generation of atomic EDM experiments that were
carried
out in the 1960s \cite{weisskopf,gould,player}.
It was realized 
quickly that the EDM \emph{enhancement factor} $R$ defined as the ratio of 
atomic EDM to the electron EDM increases with increase in the nuclear charge ($Z$) and also if there are close lying states of opposite parity since it is inversely 
proportional to the difference in their near degenerate energy levels. Hence, 
heavy rare-earth atoms with anomalously close energy levels 
of opposite parity, have large EDM enhancement factors. These factors are also
fairly large for heavy alkali atoms. 
The enhancement factor is proportional to \cite{ginges,pospelov},
\begin{eqnarray}
R \; \propto \; \frac{Z^3 \, \alpha^2}{J(J+1/2)(J+1)},
\end{eqnarray}
where $\alpha$ is the fine structure constant. This formula for order of 
magnitude estimate of EDM enhancement factors illustrates the dependence on 
nuclear charge $Z$ and the angular momentum $J$ which implies that $R$ is 
large for high $Z$ and low $J$. 

  The ground state EDMs of heavy neutral alkali atoms are of considerable 
interest to experimentalists, because: (i) the EDMs in these cases 
are several orders of magnitude larger than that of the electron,  
unlike in the hydrogen atom where the enhancement factor is large
only for the $2s$ excited state, however, it can be easily perturbed by 
external electric fields and is therefore not amenable to sensitive experiments, (ii) the availability 
of commercial lasers whose operating frequencies match with those of the 
resonant energies needed for causing transitions between the low lying 
levels and (iii) the large polarizabilities of these atoms  

The EDM of an atom or any other neutral particle is determined experimentally
by applying an external static electric field to the atom and measuring its
shift in energy that results from the interaction of the EDM with the
electric field. Consider an atom which has a permanent EDM as well as
a magnetic dipole moment. In the presence of a static electric field $\vec{E}$
and a magnetic field $\vec{B}$, the interaction Hamiltonian is given by
\begin{eqnarray}
H_{int} = - \vec{D}.\vec{E} -\vec{\mu}.\vec{B},
\end{eqnarray}
where $\vec{D}$ and $\vec{\mu}$ are, respectively, the electric and magnetic
dipole moment operators.

The application of the external fields leads to a precession of the atom.
The precession (Larmor) frequency is primarily due to the magnetic dipole
moment, but there is also a small contribution from the EDM. The observable
in an EDM experiment is the difference in the Larmor frequencies corresponding
to parallel and antiparallel configurations of $\vec{E}$ and $\vec{B}$ -
reversal of $\vec{E}$ relative to $\vec{B}$. This change in frequency is
\begin{eqnarray}
\Delta\omega_E = \frac{2DE}{\hbar}.
\end{eqnarray}
It corresponds to $\Delta\omega_E = 10^{-6} Hz$ for D$\approx 10^{-25} e-cm$ 
and $E = 10 KV/cm$.
This frequency shift corresponds to a magnetic field of $10^{-9}$G for a
diamagnetic
atom and $10^{-12}$G for a paramagnetic atom.

One of the most important systematic errors in the EDM experiment is the
magnetic field that is produced by the motion of the atoms. This field
to first order in $\frac{v}{c}$ is given in the moving frame of the atoms
as 
\begin{eqnarray}
\vec{B}_m = \frac{\vec{v}}{c}\times\vec{E}.
\end{eqnarray}
For v = 300 m/s and  E = 10 KV/cm, the motional magnetic field
$\vec{B}_m$ = 3$\times 10^{-5}$G. This magnetic field can give rise to
a frequency shift which can mimic an EDM. In the Tl experiment, two
counter-propagating beams are used to minimize this effect \cite{commins}.
EDM
experiments using optically pumped atoms in a cell have a zero average velocity
and are therefore not affected very much by the motional magnetic field
\cite{jacobs}. Both the beam and the cell experiments have their advantages and
disadvantages.
While it is possible to apply larger electric fields in the beam experiments,
the coherence times are longer in the cell experiments. The motional magnetic
fields often limit the sensitivity of the former, while leakage currents
give rise to systematic errors in the latter and they cannot be estimated easily.

EDM experiments based on laser cooled and trapped atoms in principle have the
advantages of both the beam and the cell experiments \cite{bijlsma,takahashi,romalis,chin,weiss}. In these
experiments,
one can apply large electric fields and the coherence times are long. The
leakage current problem can be overcome using a suitable configuration for
the laser trap. The systematic error due to the motional magnetic field is
virtually non-existent because of the extremely low average velocity of the
cold atoms.

The procedure of the atomic EDM measurements with cold atoms is as follows:
First, a fast atomic beam from a hot oven is slowed down by using the
Zeeman-tuning
method. After this pre-cooling stage, the atoms are trapped and cooled by a
magneto-optical trap (MOT). A high density and large number of atoms is then
loaded into the MOT within several seconds. Then the atomic beam and the
magnetic
field for the MOT is switched off, the detuning of the trapping laser is
increased
and its intensity reduced. This results in further cooling to the micro Kelvin
region by the polarization-gradient method. The next step is to perform optical
pumping to polarize the nuclear spin by the application of a circularly
polarized
resonant light pulse after completely switching off the laser fields. Finally,
a high power laser for a far blue-detuned dipole forced trap and also a high
static electric field for the EDM measurement is switched on. A probe laser
beam
will be used to measure the Larmor precession frequency. The loading and
measurement
procedure is repeated many times to reduce the statistical uncertainty.

Atomic theory is needed in combination with
experiments to extract a
variety of T violating coupling constants. Following a series of insightful
calculations
of atomic EDMs by Sandars in the 1960s and 1970s based on the relativistic
central field
potential \cite{sandars}, a number of relativistic many-body calculations on
atoms of
experimental interest have been carried out in the last two decades
\cite{das}. Ingenious experiments were initiated in the 1980s
to observe the EDMs of mercury (Hg), Cs and Tl \cite{jacobs,commins,murthy}. Subsequently systematic errors have been improved in these experiments
and today the most accurate results are available in Hg \cite{fortson} for 
diamagnetic atoms and Tl for paramagnetic atoms \cite{regan}. 
Polar molecules seem to be better candidates than heavy atoms for observing EDMs arising from the
electron EDM and there have been recent attempts to calculate the observable
in some of the  molecular EDM experiments \cite{kozlov,nayak,petrov}.
However, these calculations are in their infancy. In contrast, it is possible to perform 
very accurate calculations on alkali atoms. Therefore, to shed
light on physics beyond the SM or in fact to observe a nonzero atomic 
EDM, the heaviest of the alkali atoms, francium (Fr) seems to be 
another suitable candidate for its relatively simple spectroscopic levels and
and large enhancement factor. A number of spectroscopic analysis on Fr has been
carried out at SUNY Stony Brook using laser trapping technique \cite{gomez} 
and it has been proposed both for EDM \cite{sakemi} and atomic parity
violation (APV) \cite{stancari} measurements.
In fact, there is a preliminary calculation available for the enhancement 
factor of Fr using radial integrals for a few matrix elements
\cite{byrnes}. In the present work, we will demonstrate an approach using
the relativistic coupled-cluster method that includes Coulomb interaction
among electrons to all orders and $\mathcal{P}$ and $\mathcal{T}$ violating electron-nucleus
pseudoscalar interaction due to electron EDM up to first order but involving
contributions from continuum and all possible single and double
excited states to calculate $R$ in Fr. In this procedure, we will also demonstrate contribution from 
important intermediate states and highlight the role of various electron
correlation effects.

\section{Theoretical studies of Atomic EDM}

\subsection{General features}
\begin{figure}
\includegraphics[width=8.0cm]{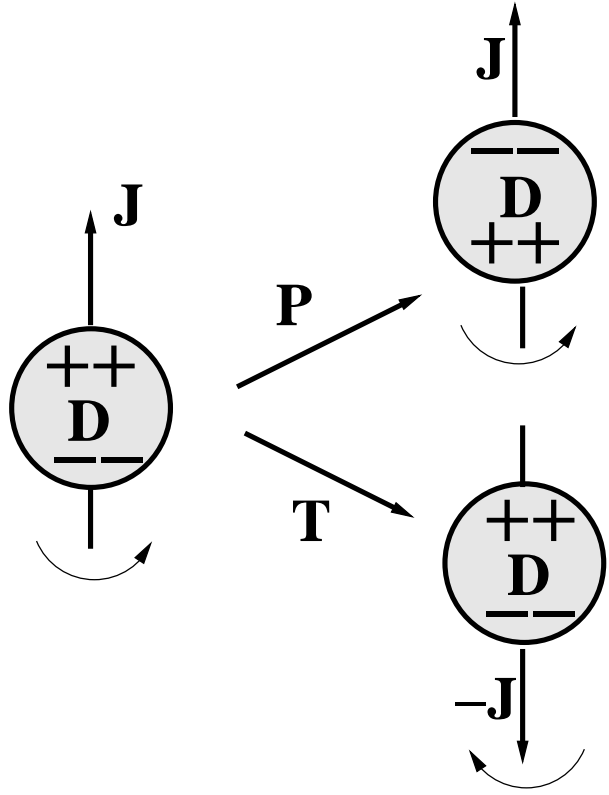}
\caption{Violation of $\mathcal{T}$ and $\mathcal{P}$ symmetries is necessary for a non-zero electric
dipole moment.}
\label{edmfig}
\end{figure}
The three discrete symmetries - charge conjugation ($\mathcal{C}$), space inversion or
parity ($\mathcal{P}$)
and time-reversal ($\mathcal{T}$) are related by the well-known CPT theorem
\cite{luders}.
According to this
theorem, while a physical system described by a local field theory may violate
any
one of these symmetries independently, it is invariant under the combined
operation
of all three of them.

The combined operation of $\mathcal{C}$ and $\mathcal{P}$ ($\mathcal{CP}$) was thought to be a good symmetry till
1964
when Christenson et al observed its violation in the decay of K$_0$ meson
\cite{christenson}.
Using
the CPT theorem, one arrives at the conclusion that this observation implies 
$\mathcal{T}$ violation. An observation of an 
electric dipole moment (EDM) of a non-degenerate physical system would be a
direct
signature of the violations of $\mathcal{T}$ as well as $\mathcal{P}$ symmetries \cite{landau}. Consider
a physical
system
with non-zero angular momentum that has a permanent EDM, $\vec{D}$. The vector $\vec{D}$
like the
magnetic dipole moment will be parallel or antiparallel to $\vec{J}$, the
angular
momentum of the system \cite{sandars1}. Under the action of $\mathcal{T}$,
$\vec{J}\rightarrow -\vec{J}$
and
$\vec{D}\rightarrow \vec{D}$. Since $\vec{D}$ is proportional to $\vec{J}$, $\mathcal{T}$
invariance
implies $\vec{D}$ = 0. If
a transformation is carried out under $\mathcal{P}$, $\vec{D}\rightarrow -\vec{D}$ but
$\vec{J}\rightarrow\vec{J}$, and this leads to $\vec{D}$ = 0 for $\mathcal{P}$ invariance.
One
therefore concludes that $\mathcal{T}$ and $\mathcal{P}$ must be violated independently for the
existence of a
non-zero EDM. The above arguments are illustrated in Fig. \ref{edmfig}.

\subsection{Atomic EDM from the electron EDM}
In a manner analogous to the anomalous magnetic moment, one can introduce an
electric dipole moment for a single electron in an electromagnetic field in a
Lorentz covariant manner into the Dirac equation. The only difference is that
it contains a pseudo-scalar Dirac operator $\gamma_5$.

The pseudo-scalar perturbation Hamiltonian for the intrinsic EDM of an electron then reads,
\begin{eqnarray}
H_{EDM} &=& i \frac{d_e}{2} \left( \bar \Psi \gamma_5 \gamma_\mu \gamma_\nu \Psi \right) F_{\mu \nu} = -\, \frac{d_e}{2} \left( \bar \Psi \gamma_5 \sigma_{\mu \nu} \Psi \right) F_{\mu \nu}
\label{hedm-ft}
\end{eqnarray}
where $d_e$ is the intrinsic EDM of the electron.

In the Pauli approximation, Eq. (\ref{hedm-ft}) reduces, in atomic units, to
\begin{eqnarray}
H_{EDM} &=& - \,d_e\, \beta \, \left( \vec \sigma \cdot \vec E \, +\, i \, \vec \alpha \cdot \vec H \right)
\label{hedm-full}
\end{eqnarray}
where $\vec \alpha, \beta$ and $\vec \sigma$ are the Dirac matrices, $\vec E$
and $\vec H$ are the total electric and magnetic fields at the site of the
electron. The second term in the rhs of Eq. (\ref{hedm-full}) is much
weaker than the first term and hence, we will consider only the
latter in the calculation of EDMs of paramagnetic atoms.
The non-relativistic limit of this interaction is $ - \,d_e\, \vec \sigma \cdot \vec E $. This form of the interaction produces a zero atomic EDM
\cite{sandars}, thus as we show below the EDM of a paramagnetic
atom due to
the intrinsic EDM of the electron is entirely a relativistic effect.
Therefore, it is necessary to use a relativistic many-body theory
to determine its size. The atomic EDM arising from the 
intrinsic EDM of the electron is discussed in detail below.

\subsection{Method of calculations}
  The total Hamiltonian for a many-electron atom, in the absence of any
external field, when the electron possesses an intrinsic EDM is given by,
\begin{eqnarray}
H = H_0 + H_{EDM}
\end{eqnarray}
where, $H_0$ is the atomic Hamiltonian given by
\begin{eqnarray}
H_0 = \sum_i \left [ c \alpha \cdot p_i + (\beta -1)c^2 + V_{nuc}(r_i) \right ] + \sum_{i \ge j} \frac{1}{r_{ij}}
\end{eqnarray}
due to the electromagnetic interaction
and $ H_{EDM}=- d_e \sum_i \beta \vec \sigma_i \cdot \vec {E}_i^{int}$ with 
the internal electric field, $\vec {E}_i^{int} = - \vec \nabla \left [ V_{nuc}(r_i) + \sum_{ i> j} \frac {1}{r_{ij}}\right ]$, exerted by the nucleus ($V_{nuc}(r_i)$)
and other electrons ($\frac{1}{r_{ij}}$).

  The application of external electric field also induces EDM to the atom.
Hence, the total perturbed Hamiltonian $H^{(1)}$ in the presence of
an external electric field is given by,
\begin{eqnarray}
H^{(1)} \, =\, H_{EDM}-\, d_e \sum_i \beta \vec \sigma_i \cdot \vec E\, - \, e \sum_i \vec r_i \cdot \vec E
\end{eqnarray}

In the time-independent perturbation theory, the first order shift in energy is given by,
\begin{eqnarray}
E_m^{(1)}\, &=& \,\langle \Psi_m^{(0)} \vert\, H^{(1)}\, \vert \Psi_m^{(0)}\rangle \nonumber \\
            &=& \, - \sum_i \left [ \langle \Psi_m^{(0)} \vert\, d_e \, \beta \vec \sigma_i \cdot \vec {E}_i^{int} \vert \Psi_m^{(0)}\rangle \, -\, \langle \Psi_m^{(0)} \vert\, d_e \beta \vec \sigma_i \cdot \vec E\, \vert \Psi_m^{(0)}\rangle - \langle \Psi_m^{(0)} \vert\, e \vec r_i \cdot \vec E \vert \Psi_m^{(0)}\rangle \right ]
\end{eqnarray}

Assuming that the applied field is in the positive z-direction,
\begin{eqnarray}
E_m^{(1)}\,  = \, \sum_j \left [ -d_e \langle \Psi_m^{(0)} \vert\, \beta \vec \sigma_{z,j} \cdot \vec {E}_j^{int} \vert \Psi_m^{(0)}\rangle \, - \, d_e \langle \Psi_m^{(0)} \vert\, \beta \vec \sigma_{z,j} \vert \Psi_m^{(0)}\rangle \vert \vec E \vert  - \langle \Psi_m^{(0)} \vert\, e \vec z_j \vert \Psi_m^{(0)}\rangle \vert \vec E \vert \right ]
\end{eqnarray}

Noting the fact that, the operators in the first and third terms are
\emph {odd} under parity and the expectation value of \emph {odd parity}
operator vanishes in a state of definite parity, the only non-vanishing term
(corresponds to the \emph {even parity} operator) is given by,
\begin{eqnarray}
E_m^{(1)}\, =\, - d_e\, \langle\, \Psi_m^{(0)}\, \vert\, \beta \sum_j \vec \sigma_{z,j} \, \vert\, \Psi_m^{(0)}\rangle \vert \vec E \vert
\label{em1}
\end{eqnarray}

As the strength of the perturbation is sufficiently weak, we consider only up
to the first-order perturbation in the  wavefunction. The first-order
perturbed wavefunction $\vert \Psi_m^{(1)}\rangle$ is given by,
\begin{eqnarray}
\vert\, \Psi_m^{(1)}\,\rangle \,=\, \sum_{n \neq m } \,\frac{\vert\, \Psi_n^{(0)} \rangle \langle\, \Psi_n^{(0)}\, \vert\, H^{(1)} \,\vert\, \Psi_m^{(0)}\rangle}{E_m^{(0)}\,-\,E_n^{(0)}}
\end{eqnarray}

The second order shift in energy due to EDM as a perturbation is given by,
\begin{eqnarray}
E_m^{(2)}\, &=& \,\langle\, \Psi_m^{(0)} \,\vert\, H^{(1)}\, \vert\, \Psi_m^{(1)}\rangle \nonumber \\
\nonumber \\
 &=&\, \sum_{n \neq m} \,\frac{ \langle\, \Psi_m^{(0)}\,\vert\, H^{(1)} \, \vert \,\Psi_n^{(0)}\rangle \langle \,\Psi_n^{(0)}\,\vert\, e\, \sum_j \vec z_j \,\vert\, \Psi_m^{(0)}\rangle} {E_m^{(0)}\,-\,E_n^{(0)}}
\end{eqnarray}

On inserting the expression for $H^{(1)}$ in the above equation and discarding
the terms containing ${d_e}^2$ and $E^2$ while expanding, we get those terms
which are first order in perturbation and linearly proportional to the applied
uniform electric field as below;
\begin{eqnarray}
E_m^{(2)}\, &=& \: \bigg\{ \sum_{n \neq m} \,\frac{ \langle\, \Psi_m^{(0)}\,\vert\, H_{EDM} \, \vert \,\Psi_n^{(0)}\rangle \langle \,\Psi_n^{(0)}\,\vert\, e\, \sum_j \vec z_j \,\vert\, \Psi_m^{(0)}\rangle} {E_m^{(0)}\,-\,E_n^{(0)}}  \nonumber \\
&+&  \sum_{n \neq m} \,\frac{ \langle \,\Psi_n^{(0)}\,\vert\, e\, \sum_j \vec z_j \,\vert\, \Psi_m^{(0)}\rangle \langle\, \Psi_m^{(0)}\,\vert\, H_{EDM} \, \vert \,\Psi_n^{(0)}\rangle}{E_m^{(0)}\,-\,E_n^{(0)}} \bigg\} \vert \vec E \vert
\label{em2}
\end{eqnarray}

Thus, the total shift in energy would be,
\begin{eqnarray}
E_m\, &=& \, E_m^{(1)}\, +\, E_m^{(2)} \nonumber \\
&=& \: \bigg\{\,- d_e \langle\, \Psi_m^{(0)}\, \vert\, \beta \sum_j \vec \sigma_{z,j}\, \vert\, \Psi_m^{(0)}\rangle  \,
+\,  \sum_{n \neq m} \,\frac{ \langle\, \Psi_m^{(0)}\,\vert\, H_{EDM} \, \vert \,\Psi_n^{(0)}\rangle \langle \,\Psi_n^{(0)}\,\vert\, e\, \sum_j \vec z_j \,\vert\, \Psi_m^{(0)}\rangle} {E_m^{(0)}\,-\,E_n^{(0)}}  \nonumber \\
&+&  \sum_{n \neq m} \,\frac{ \langle \,\Psi_n^{(0)}\,\vert\, e\, \sum_j \vec z_j \,\vert\, \Psi_m^{(0)}\rangle \langle\, \Psi_m^{(0)}\,\vert\, H_{EDM} \, \vert \,\Psi_n^{(0)}\rangle}{E_m^{(0)}\,-\,E_n^{(0)}} \bigg\} \vert \vec E \vert
\end{eqnarray}

The coefficient of the linear shift in energy due to the applied electric field is the total electric dipole moment of an atom i.e, $\varepsilon = - \vec D \cdot \vec E$. Thus, the total EDM of an atom is given by,
\begin{eqnarray}
\langle D \rangle &=&  \: \bigg\{\, d_e \langle\, \Psi_m^{(0)}\, \vert\, \beta \sum_j \vec \sigma_{z,j} \, \vert\, \Psi_m^{(0)}\rangle \, -\, \sum_{n \neq m} \,\frac{ \langle\, \Psi_m^{(0)}\,\vert\, H_{EDM} \, \vert \,\Psi_n^{(0)}\rangle \langle \,\Psi_n^{(0)}\,\vert\, e\, \sum_j \vec z_j \,\vert\, \Psi_m^{(0)}\rangle} {E_m^{(0)}\,-\,E_n^{(0)}}  \nonumber \\
&-& \sum_{n \neq m} \,\frac{ \langle \,\Psi_n^{(0)}\,\vert\, e\, \vec z \,\vert\, \Psi_m^{(0)}\rangle \langle\, \Psi_m^{(0)}\,\vert\, H_{EDM} \, \vert \,\Psi_n^{(0)}\rangle}{E_m^{(0)}\,-\,E_n^{(0)}} \bigg\}
\label{total-edm}
\end{eqnarray}

The total atomic EDM given by Eq. (\ref{total-edm}) can be further simplified to the effective one-electron form given by,
\begin{eqnarray}
\langle D \rangle \, =\, \sum_j \left [ \frac{2 i \, c\, d_e}{\hbar}\, \bigg\{\sum_{n \neq m} \frac{ \langle\, \Psi_m^{(0)}\,\vert\, \beta \, \gamma^5\, {p}_j^2 \,\vert\, \Psi_n^{(0)}\,\rangle \langle\, \Psi_n^{(0)}\,\vert\, \vec z_j \,\vert\, \Psi_m^{(0)}\,\rangle} {E_m^{(0)}\,-\,E_n^{(0)}} \bigg\}\, \right ] + h.c.
\label{d-eff}
\end{eqnarray}

Here, $\beta$ and $\gamma^5$ are the Dirac matrices, $\vec p$ is the 3-momentum, $\hbar \left( = \frac{h}{2\pi} \right)$ is the modified Planck's constant, $E_m^{(0)}$ and $E_n^{(0)}$ are the zeroth order energies of the states $m$ and $n$ respectively, and the abbreviation $h.c.$ stands for hermitian conjugate.
Eq. (\ref{d-eff}) shows that $H_{EDM}$ in Eq. (\ref{total-edm}) reduces to
an effective one-body term given by $H_{EDM}^{eff}=2 i \, c\, d_e \sum_j \beta \gamma^5 {p}_j^2$.

$H_{EDM}^{eff}$ can be responsible for mixing atomic states of opposite parities.
Its strength is sufficiently weak for it to be considered as a first-order
perturbation. It is, therefore, possible to write the {\it m'th} state atomic
wavefunction as
\begin{equation}
|\Psi_m \rangle = |\Psi_m^{(0)} \rangle + d_e |\Psi_m^{(1)} \rangle .
\label{eqn17}
\end{equation}

Therefore, we have
\begin{eqnarray}
 \langle D \rangle &=& d_e \frac {\langle \Psi_m^{(0)}| D |\Psi_m^{(1)} \rangle + \langle \Psi_m^{(1)}| D |\Psi_m^{(0)} \rangle } {\langle \Psi_m^{(0)}|\Psi_m^{(0)} \rangle } ,
\label{eqn18}
\end{eqnarray}
where D is the electric dipole (E1) operator.

The above first order perturbed wave functions due to $H_{EDM}^{eff}$
can be calculated by summing over a few important intermediate states. 
However, the accuracy of this approach is rather limited. We have
developed an approach based on the relativistic coupled-cluster (RCC) theory
that can overcome the sum-over-states approach by directly solving the 
first order perturbed equation 
\begin{eqnarray}
(H^{(0)} - E_m^{(0)})|\Psi_m^{(1)}\rangle = (E_m^{(1)} - H_{EDM}^{eff}) |\Psi_m^{(0)}\rangle , \label{eqn20}
\end{eqnarray}
where $E_m^{(1)}$ vanishes since $H_{EDM}^{eff}$ is an odd parity operator. We 
present below the formulation of this problem based on RCC theory.

\subsection{RCC theory of atomic EDM}
Using RCC theory, the atomic wavefunction $|\Psi_m^{(0)} \rangle$ for a single valence ($m$) open-shell system is given by \cite{lindgren,mukherjee}
\begin{eqnarray}
|\Psi_m^{(0)} \rangle = e^{T^{(0)}} \{1+S_m^{(0)}\} |\Phi_m \rangle , \label{eqn7}
\label{eqn35}
\end{eqnarray}
where we define $|\Phi_m \rangle= a_m^{\dagger}|\Phi_0\rangle$, with $|\Phi_0\rangle$ as the Dirac-Fock (DF) state for the closed-shell system. The curly
bracket in the above expression represents normal order form.

In the single and double excitations approximation coupled-cluster (CCSD) method, we have
\begin{eqnarray}
T^{(0)} &=& T_1^{(0)} + T_2^{(0)} , \nonumber \\
S_m^{(0)} &=&  S_{1m}^{(0)} + S_{2m}^{(0)} , \label{eqn36}
\end{eqnarray}
\noindent
where $T_1^{(0)}$ and $T_2^{(0)}$ are the single and double particle-hole
excitation operators for the core electrons and $S_{1m}^{(0)}$ and
$S_{2m}^{(0)}$ are the single and double excitation operators for the valence
electron, respectively. The amplitudes
corresponding to these operators can be determined by solving the relativistic
coupled-cluster singles and doubles equations.

These amplitudes are obtained by solving the following equations
\begin{eqnarray}
&&\langle \Phi^L |\{\widehat{He^T}\}|\Phi_0 \rangle = \delta_{0,L} \Delta E_{corr} \label{eqn9}\\
&&\langle \Phi_m^L|\{\widehat{He^T}\}S_m|\Phi_m\rangle = - \langle \Phi_m^L|\{\widehat{He^T}\}|\Phi_m\rangle +  \langle \Phi_m^L|1+S_m|\Phi_m\rangle \langle \Phi_m|\{\widehat{He^T}\} \{1+S_m\} |\Phi_v\rangle \nonumber \\
&& \ \ \ \ \ \ \ \ \ \ \ \ \ \ \ \ \ \ \ \ \ \ \ \ \ = -  \langle \Phi_m^L|\{\widehat{He^T}\}|\Phi_m\rangle - \langle \Phi_m^L|\delta_{L,m}+S_m|\Phi_m\rangle \text{IP}, \label{eqn37}
\end{eqnarray}
where the superscript $L (=1,2)$ represents excited states from the
corresponding DF states, the wide hat symbol denotes connected terms,
$\Delta E_{cor}$ and IP represents the correlation energy and ionization
potential of the valence electron $'m'$, respectively. We consider the
Dirac-Coulomb (DC) Hamiltonian in our calculation given by
\begin{eqnarray}
H &= & [H_0] + [V_{es}] \nonumber \\
  & = & \sum_i^N \Lambda_i^+ \left [ c \alpha \cdot p_i + (\beta -1)
c^2 + V_{nuc}(r_i) +U(r_i) \right ] \Lambda_i^+ \nonumber \\
 && + [ \sum_{i>j}^N \Lambda_i^+ \Lambda_j^+ \frac {1}{r_{ij}} \Lambda_j^+ \Lambda_i^+ - \sum_i^N \Lambda_i^+ U(r_i) \Lambda_i^+ ], \label{eqn38}
\end{eqnarray}
where $H_0$ is the DF Hamiltonian, $V_{es}$ is the Coulomb residual term in
atomic units that is neglected in the DF calculation
and $\Lambda^+$ are the projection operators on to the positive-energy states
of the Dirac Hamiltonian in the nuclear ($V_{nuc}(r_i)$) and DF ($U(r_i)$)
potentials.

The most important triple excitations have been considered by constructing
triple excitation operators \cite{kaldor}
\begin{equation}
S_{mbc}^{pqr(0)}\ =\ \frac{\widehat{V_{es}T_2^{(0)}}+\widehat{V_{es}S_{m2}^{(0)}}}{\epsilon_b+\epsilon_c-\epsilon_q-\epsilon_r}, \label{eqn39}
\end{equation}
where $\epsilon$'s are the orbital energies. The above operators are used to
construct single and double open-shell cluster amplitudes by connecting
further with the CCSD operators and evaluating contributions to IP which are
also involved in the amplitude determining equations. Finally, these amplitudes
are solved self-consistently.

In the presence of the EDM, the exact atomic wavefunction can be written as
\begin{eqnarray}
|\Psi_m \rangle = e^T \{1+S_m\} |\Phi_m \rangle, \label{eqn40}
\end{eqnarray}
where the cluster amplitudes are given by
\begin{eqnarray}
T = T^{(0)} + d_e T^{(1)} , \nonumber \\
S_m =  S_m^{(0)} + d_e S_m^{(1)} . \label{eqn41}
\end{eqnarray}
$T^{(1)}$ and $S_m^{(1)}$ are the first order in $d_e$ corrections to the
cluster operators  $T^{(0)}$ and $S_m^{(0)}$, respectively. In the CCSD
method, we have
\begin{eqnarray}
T^{(1)} &=& T_1^{(1)} + T_2^{(1)} , \nonumber \\
S_m^{(1)} &=&  S_{1m}^{(1)} + S_{2m}^{(1)} . 
\end{eqnarray}

The amplitudes of these operators are solved, keeping up to linear in $d_e$, by the following equations
\noindent
\begin{eqnarray}
&&\langle \Phi^L | \widehat{\overline{H_N^{(0)}} T^{(1)}} + \overline{H_{EDM}^{eff}}| \Phi_0 \rangle = 0 , \label{eqn42} \\
&&\langle\Phi_m^L| \widehat{\overline{H_N^{(0)}} S_m^{(1)}} +\widehat{\overline{H_N^{(0)}}T^{(1)}} + \widehat{\overline{H_N^{(0)}}T^{(1)} S_m^{(0)}} +\overline{H_{EDM}^{eff}} + \widehat{\overline{H_{EDM}^{eff}}S_m^{(0)}}|\Phi_m\rangle =  - \langle\Phi_m^L|S_m^{(1)}|\Phi_m\rangle \text{IP} , \label{eqn43}
\end{eqnarray}
\noindent
where Hamiltonian operators with overline as defined as $\overline{H}_N =e^{-T^{(0)}}\{H\}e^{T^{(0)}}$. These are computed after determining $T^{(0)}$.

After solving the CCSD amplitudes we determine $\langle D \rangle$ by
\begin{eqnarray}
R = \frac {\langle D \rangle}{d_e} = \frac {<\Phi_m |\{ S_m^{(1)^{\dagger}} + (1+T^{(1)^{\dagger}}) S_m^{(0)^{\dagger}} \} e^{T^{(0)^{\dagger}}} D e^{T^{(0)}} \{ (1+ T^{(1)} ) S_m^{(0)} + S_m^{(1)} \} |\Phi_m > }
{(1+N_m^{(0)})} . \label{eqn44}
\end{eqnarray}
In the above expression, we define $\overline{D^{(0)}} = e^{T^{{(0)}^\dagger}}De^{T^{(0)}}$ and $N_m^{(0)} = \langle \Phi_m| S_m^{{(0)}^{\dagger}}e^{T^{{(0)}^\dagger}}e^{T^{(0)}}S_m^{(0)} |\Phi_m \rangle$ for the valence electron '$m$'. These terms are evaluated using the generalized Wick's theorem
\cite{lindgren} by constructing effective one-body, two-body etc. terms.

 As explained in our earlier work on EDM arising from scalar-pseduoscalar 
interaction in Cs and Tl \cite{sahoo}, the important RCC terms in the
above expression are $\overline{D^{(0)}} T_1^{(1)}$, $\overline{D^{(0)}} S_{1m}^{(1)}$ and $\overline{D^{(0)}} S_{2m}^{(1)}$. The first 
term corresponds to the core-correlation effects whereas the second and third 
terms correspond to the valence correlation effects. Core-polarization and
pair-correlation effects arising through the singly excited states
are considered through $\overline{D^{(0)}} S_{1m}^{(1)}$ and
important core-polarization effects from the doubly excited states 
are taken into account through $\overline{D^{(0)}} S_{2m}^{(1)}$.
Important excited states that contribute significantly through these
terms are given below.

\section{Construction of basis functions}
The EDM interactions for both closed and open shell atoms are sensitive to the
nuclear region. Therefore, the Gaussian type orbitals (GTOs) that produce good wave functions in the 
nuclear region can be used to calculate accurate $\langle D \rangle$. For 
atomic wavefunction calculations, they are given by \cite{rajat}
\begin{equation}
F_{i,k}(r) = r^k e^{-\alpha_ir^2} , \label{eqn45}
\end{equation}
where $k=0,1,..$ for s,p,.. type orbital symmetries, respectively. For the
exponents, we have used
\begin{equation}
\alpha_i = \alpha_0 \beta^{i-1} . \label{eqn46}
\end{equation}

We consider excitations from all occupied orbitals (holes)in both the DF and RCC calculations.
The orbitals are generated on a grid as in the numerical code GRASP \cite{parpia1}. The finite size of the nucleus has been accounted for by considering a two-parameter Fermi nuclear charge distribution approximation given by
\begin{equation}
\rho = \frac {\rho_0} {1 + e^{(r-c)/a}}. \label{eqn47}
\end{equation}
We use values of $\rho_0$, $c$ and $a$ as given by Parpia and Mohanty \cite{parpia}.

\section{Results and Discussions}
\begin{table}[h]
\caption{Contributions from important RCC terms to the $R=D_a/d_e$ calculations 
of the ground state in Fr. $c.c$ represents conjugate terms. Here contributions
given by $Norm$ and $Others$ correspond to normalization corrections higher
order terms those are not mentioned in this table, respectively.} 
\begin{ruledtabular}
\begin{tabular}{ll}
RCC terms & $R=D_a/d_e$ \\
\hline
 &  \\
Dirac-Fock (core) & \hfill$25.77$ \\
Dirac-Fock (virtual) & \hfill$695.44$ \\
\hline
 &  \\
$D T_1^{(1)}+c.c. $ & \hfill$43.39$ \\
$\overline{D^{(0)}} S_{1m}^{(1)}+c.c.$ & \hfill$1000.19$ \\
$\overline{D^{(0)}} S_{2m}^{(1)} $ & \hfill$-64.94$ \\
$S_{1m}^{(0)\dagger} \overline{D^{(0)}} S_{1m}^{(1)} +c.c.$ & \hfill$-18.07$ \\
$S_{2m}^{(0)\dagger} \overline{D^{(0)}} S_{1m}^{(1)} +c.c.$ & \hfill$-59.18$ \\
$S_{1m}^{(0)\dagger} \overline{D^{(0)}} S_{2m}^{(1)} +c.c.$ & \hfill$-2.80$ \\
$S_{2m}^{(0)\dagger} \overline{D^{(0)}} S_{2m}^{(1)} +c.c.$ & \hfill$19.26$ \\
$Norm$ & \hfill$-24.42$  \\
$Others$ & \hfill$1.51$  \\
 &  \\
Total & \hfill$894.93$ \\
\end{tabular}
\end{ruledtabular}
\label{tab1}
\end{table}

In Table \ref{tab1}, we present our enhancement factor calculations to the
ground state of Fr. We obtain the final result $R=894.93$ which agrees fairly 
well with Byrnes et al. who have obtained $910(46)$
 (error bar is quoted as $\sim 5\%$) \cite{byrnes}. However, this is just a coincidence. Byrnes et al. have used a sum-over-states approach and have only
considered singly excited valence states as the intermediate states. They have 
therefore not taken into account pure core correlation
and contributions from the doubly excited states. In their calculation,
the $7s_{1/2}-7p_{1/2}$ E1 matrix element was taken from 
experiment and only $7p_{1/2}-10p_{1/2}$ discrete states were used in the 
calculation and the contributions of the continuum states were included in an 
approximate way. Another limitation in the calculation of Byrnes et al is in 
their treatment of the internal electric field in their parity and 
time-reversal violating Hamiltonian. They have considered only the potential 
that an electron sees from the nucleus but not the other electrons of the 
atom. We have included 
contributions from all the core-electrons and all possible single and double
excited states through the RCC method in our calculation.

 We have also explicitly given the DF contributions from core and virtual 
orbitals in Table \ref{tab1}. It is clear from this that the contributions
due to core orbitals are not small in the present system. Our $D T_1^{(1)}+c.c. $ RCC terms at the lowest order correspond to the above DF core contributions. 
Comparing contributions at the DF and the RCC levels, it shows that the
all order core correlation effects are almost twice as much as the DF
contribution. Similarly, our $\overline{D^{(0)}} S_{1m}^{(1)}+c.c.$
RCC terms at the lowest order corresponds to the DF results only due to the
virtual orbitals. From Table \ref{tab1}, it is also clear that the 
all order correlation effects are significant. 

\begin{table}[h]
\caption{Break down of $D T_1^{(1)}+c.c. $ contributions using DF reduced E1 matrix elements from core orbitals.}
\begin{center}
\begin{tabular}{lccc}
\hline
\hline
 & $\langle 7s_{1/2}||D||np_{1/2}\rangle_{DF}$ & $\langle np_{1/2}|T_1^{(1)}|7s_{1/2}\rangle$ & $R$ \\
\hline
 & & \\
 $n=2$ & $-0.15\times 10^{-3}$ & 1.55 & $0.19\times 10^{-3}$\\
 $n=3$ & $0.88\times 10^{-3}$ & $-3.35$ & $0.24\times 10^{-2}$\\
 $n=4$ & $-0.52\times 10^{-2}$ & 7.56 & $0.32\times 10^{-1}$\\
 $n=5$ & $0.38\times 10^{-1}$ & $-20.31$ & 0.63 \\
 $n=6$ & 0.54 & $-100.01$ & 44.16 \\
 & & \\
\hline
\hline
\end{tabular}
\end{center}
\label{tab2}
\end{table}
  Contributions from the doubly excited states arise through the $S_{2m}^{(1)}$
operator. It can be seen from the above table that these states contribute to
about 7\% iof the final result, but with opposite sign.  Therefore, the 
error bar quoted (which is $\sim 5\%$) by Byrnes et al. without considering 
the doubly excited states contributions does not seem to be correct.  
Again, the normalization correction ($Norm$) in our calculation
is also about 3\%, which is a significant contribution, and this appears to be
missing in the work of Byrnes et al.

  The trends exhibited by the correlation effects in the present work are similar to those in the case of rubidium (Rb)
\cite{nataraj}, but the amount of core correlation 
effects are substantially larger  for Fr. Interestingly, we observe that the 
correlation contributions for Rb \cite{nataraj}, Cs \cite{nataraj} and Fr
are about 24\%, 22\% and 20\%, respectively. This is because of the fact 
that as the size of the system increases contributions from the 
doubly excited states increases with opposite sign and hence there are 
strong cancellations in the heavy systems.

\begin{table}[h]
\caption{Break down of $D S_{1m}^{(1)}+c.c. $ contributions using DF reduced E1 matrix elements from important virtual orbitals.}
\begin{center}
\begin{tabular}{lccc}
\hline
\hline
 & $\langle 7s_{1/2}||D||np_{1/2}\rangle_{DF}$ & $\langle np_{1/2}|S_{1m}^{(1)}|7s_{1/2}\rangle$ & $R$ \\
\hline
 & & \\
 $n=7$ & 5.12 & 227.26 & 949.69 \\
 $n=8$ & $-0.57$ & $-81.71$ & 37.79\\
 $n=9$ & 0.18 & 115.55 & 16.66 \\
 $n=10$ & $0.17\times 10^{-1}$ & $-91.63$ & $-1.25$ \\
 $n=11$ & $-0.53\times 10^{-1}$ & $45.35$ & $-1.95$ \\
 $n=12$ & $0.17\times 10^{-1}$ & $-17.37$ & $-0.24$ \\
 $n=13$ & $-0.21\times 10^{-2}$ & $5.17$ & $-0.87\times 10^{-2}$ \\
 $n=14$ & $-0.29\times 10^{-3}$ & $2.02$ & $-0.48\times 10^{-3}$ \\
 & & \\
\hline
\hline
\end{tabular}
\end{center}
\label{tab3}
\end{table}

 In Table \ref{tab2}, we present the core orbital contributions combining 
the reduced matrix element of E1 matrix element obtained using DF method
and RCC amplitude obtained using $T_1^{(1)}$ operator. This shows that
the upper most $p_{1/2}$ core orbital, i.e. $6p_{1/2}$ orbital, contributes
almost through this term and the remaining core orbital contributions
are small. This can be understood as the energy difference between the 
valence orbital, $7s_{1/2}$ and $6p1/2$ core orbital is small and hence
it contributes large.
\begin{table}[h]
\caption{Break down of $D S_{2m}^{(1)}+c.c. $ contributions using DF reduced E1 matrix elements from important doubly excited states. Here indices $m$, $l$ and $k$ represent valence, core and virtual orbitals, respectively.}
\begin{center}
\begin{tabular}{lccc}
\hline
\hline
 & $\langle l||D||k \rangle_{DF}$ & $\langle m l |S_{2m}^{(1)}| m k \rangle$ & $R$ \\
\hline
 & & \\
 $l=6p1/2$; $k=7s_{1/2}$ & 0.54 & 11.05 & $-5.25$ \\
 $l=6p1/2$; $k=10s_{1/2}$ & 0.54 & $13.78$ & $-2.22$ \\
 $l=6p1/2$; $k=9d_{3/2}$ & $-1.10$ & $-11.95$ & $-2.22$ \\
 $l=6p1/2$; $k=10d_{3/2}$ & $-1.48$ & $14.89$ & $-4.93$ \\
 $l=6p3/2$; $k=10d_{3/2}$ & $-0.74$ & $-27.17$ & $-4.09$ \\
 $l=6p3/2$; $k=7d_{5/2}$ & $1.78$ & $-25.17$ & $-10.24$ \\
 $l=6p3/2$; $k=8d_{5/2}$ & $-1.18$ & $16.31$ & $-4.48$ \\
 $l=6p3/2$; $k=9d_{5/2}$ & $-1.80$ & $24.87$ & $-10.55$ \\
 $l=6p3/2$; $k=10d_{5/2}$ & $-2.21$ & $27.31$ & $-14.80$ \\
 $l=6p3/2$; $k=11d_{5/2}$ & $1.10$ & $-7.85$ & $-2.50$ \\
 & & \\
\hline
\hline
\end{tabular}
\end{center}
\label{tab4}
\end{table}

 As in the case of the core orbital contributions, we have also investigated the
role of various virtual orbitals considering the most dominant operator,
$S_{2m}^{(1)}$, and E1 matrix element obtained using DF method. These results
are reported in Table \ref{tab3}. As seen in this table, $7p_{1/2}$,
$8p_{1/2}$ and $9p_{1/2}$ orbitals contribute the most. In our DF calculations, the
orbitals up to $8p_{1/2}$ are bound and $9p_{1/2}$ orbital onwards as continuum orbitals, respectively.
Large contribution from the $7p_{1/2}$ orbital can be understood based on the
fact that the energy difference between the valence orbital 
$7s_{1/2}$ and virtual orbital $7p_{1/2}$ is very small and the overlap of these two orbitals s large at distances close to and far away from the nucleus. Since 
the density of the continuum orbital $9p_{1/2}$ in the nuclear region is large,  it also gives a large contribution.

 We have also determined the contributions from the most important 
doubly excited states by combining the $S_{2m}^{(1)}$ operator with the E1
matrix elements obtained using the DF method and given them in Table
\ref{tab4}. As a special condition, our $S_{2m}^{(1)}$ can excite the
valence orbital $7s_{1/2}$ to itself (a spectator), but other virtuals may
go to the core orbitals by the dipole operator. This is a special type
of core-polarization effect and which manifests through doubly excited states. From 
the above table, we find that virtual orbitals from the $d$ symmetry
are contribute significantly through this process.

\section{Conclusion}
We have discussed the relativistic theory of the electric dipole moments of  
paramagnetic atoms arising from the electric dipole moment of the electron. We have employed the relativistic
coupled-cluster method to calculate the atomic wavefunctions to all orders in 
Coulomb interaction and one order in the $\mathcal{P}$ and $\mathcal{T}$ violating interaction and
obtained the EDM enhancement factor for francium; the heaviest of all the
alkali atoms. Our result has been 
compared with the available semi-empirical result and contributions from
various correlation effects and important intermediate states have been
explicitly given.
 
\section{Acknowledgment}
{\it This paper is dedicated in honor of Russel Pitzer, a great pioneer in the developments and applications of
relativistic quantum chemistry}.
BKS thanks NWO for financial support through the VENI fellowship with grant
680-47-128. DM thanks DST (New Delhi) for the award of the J. C. Bose
Fellowship and the Jahawarlal Center for Advanced Scientific Research, Bangalore
for conferring on him an honorary professorship. The computations
were performed using the Tera-flop Super computer, Param Padma in C-DAC, Bangalore.

\end{document}